\documentstyle[12pt]{article}
\begin{document}
\input epsf

\begin{center}
\large{{\bf Study of the Magnitude-Redshift Relation for type Ia Supernovae
in a Model resulting from a Ricci-Symmetry}}
\end{center}

\vspace{.5cm}  
\begin{center}
 R. G. Vishwakarma\footnote{IUCAA, Post Bag 4, Ganeshkhind, Pune 411 007, 
 India \\E-mail: vishwa@iucaa.ernet.in}
\end{center}  

\vspace{1cm}
 
\noindent
{\bf Abstract:}

\noindent
Models with a dynamic cosmological term $\Lambda~ (t)$ are becoming
popular as they solve 
the cosmological constant problem in a natural way. Instead of 
considering any ad-hoc assumption for the variation of $\Lambda$, we
consider a particular symmetry, the contracted Ricci-collineation
along the fluid flow, in Einstein's theory.
We show that apart from having interesting properties, this symmetry 
does demand $\Lambda$ to be a function of the scale factor of the
Robertson-Walker metric.
In order to test the consistency of the resulting model with observations, 
we study the magnitude-redshift relation for the type Ia supernovae data
from Perlmutter et al. The data fit the model very well and require a positive
non-zero $\Lambda$ and a negative deceleration parameter. The best-fitting
flat model is obtained as $\Omega_0 \approx 0.5$ with $q_0 \approx -0.2$.

\vspace{.5cm}


\vspace{1cm}

\noindent
{\bf 1. INTRODUCTION}
\vspace{.2cm}

\noindent
The recent measurements of the CMB anisotropy [1] and the observations of type 
Ia supernovae made, independently, by
Perlmutter et al [2] and Riess et al [3] consistently demand
a significant and positive  cosmological constant $\Lambda$.
These observations suggest Friedmann models with negative pressure-matter 
in which
the expansion of the universe is accelerating.
Observations of gravitational lensing [4]
also  indicate the presence of a non-zero $\Lambda$.

On the other hand, a dynamical $\Lambda (t)$ has been considered in numerous
papers in order to explain its observed small value which is about
120 orders of magnitude below the value for the vacuum energy density
predicted by quantum field theory- the so-called cosmological constant
problem [5]. This phenomenological solution is based on the argument
that $\Lambda$ relaxed to its present estimate due to the expansion of
the universe. (It is customary to associate a positive cosmological
constant $\Lambda$ with a vacuum energy density
$\rho_{\mbox{{\scriptsize v}}} \equiv \Lambda/8\pi G$.)

As the dynamics of the variable $\Lambda$-models depends sensitively on the
chosen dynamic law for the variation of $\Lambda$ and, in general, becomes
altogether
different from the dynamics of the corresponding constant $\Lambda$-models,
there is no reason to believe that the observations of distant objects
would also agree with the  variable $\Lambda$-models, given that they agree
with the corresponding constant $\Lambda$-ones, especially for the same
estimates of the parameters.
In this view, it would be worth while to test the consistency of the
above-mentioned observations with the variable $\Lambda$-models which
solve the cosmological constant problem in a natural way.
It may be mentioned that the quintessence or the exotic `x-fluid' models
($-1<p_{\rm x}/\rho_{\rm x}<0$) are also capable to explain, in a natural way, the
smallness of the present vacuum energy density, since the associated vacuum
energy density dynamically evolves towards zero with the expansion of the
universe. However, as Garnavich et al [6] have shown, if one attempts to
constrain the equation of state of the dark energy component (that may 
have contributed to accelerating the cosmic expansion), with the Ia 
supernova data, one finds that the data is consistent with either a 
cosmological constant, or a scalar field that has, on average, an 
equation-of-state parameter similar to the one for the cosmological constant, 
i.e., -1. However, we note that $p_{\rm x}/\rho_{\rm x} =-1$ corresponds to the
cosmological  $\Lambda$ which, in view of the conservation of the energy
momentum tensor of the  exotic component, becomes a true constant and, hence,
cannot solve the  cosmological constant problem. However, as we shall see
later, the usual kinematical $\Lambda$ can always be, in general, a function
of time as the  conserved quantity in this case is matter plus vacuum, and not
the vacuum only [7].

We note that a number of models with $\Lambda$ as a function of time have
been presented in recent years [8, 10-13]. 
Different phenomenological laws, for
the decay of $\Lambda$, have been proposed in these models which are
either from dimensional arguments or from ad-hoc assumptions.
Though the precise mechanism of vacuum-decay, which should come from the
fundamental interactions, is not yet known, it would, however, be worth while
to look for some symmetry principles which actually demand the variation of 
$\Lambda$. Moreover, it is always reasonable to consider symmetry
properties of spacetime rather than considering ad-hoc assumptions for
the variation of $\Lambda$.

In this paper, we consider, on the level of classical general relativity,
a particular Ricci-symmetry which is the contracted Ricci-collineation along
the fluid flow vector and show that, apart from having interesting 
properties, this symmetry does demand $\Lambda$ to be a function of time 
(and space, in general). This is done in section 2. In section 3, we derive
the magnitude ($m$) - redshift ($z$) relation in the resulting model to test 
its consistency with the Ia supernovae data from Perlmutter et al [2]. 
The model fits the data very well. The numerical results are discussed in 
section 4 followed by a concluding section.

Earlier some cosmological models with this symmetry were discussed 
 by considering  particular sets of initial and boundary
conditions and by assuming that $\Lambda=\Lambda(t)$ in some of them [8-10]. 
However, as we shall see in section 2,
the incorporation of this particular symmetry in the Einstein theory
does demand $\Lambda$ to be a variable and there is no need to assume
$\Lambda=\Lambda(t)$ a priori. For the ready reference and
completeness, we reproduce the model in the following.

\vspace{1cm}
\noindent
{\bf 2. SYMMETRY CONSIDERATIONS AND THE RESULTING MODEL}

\vspace{.2cm}

We begin our discussion by considering the contracted
Ricci-collineation. The motivation for considering this particular
symmetry
will be discussed later. We note that a spacetime is said to admit a
Ricci-collineation along a field vector $\eta^i$ if [14]
\begin{equation}
{\cal L}_\eta R_{ij}=0,
\end{equation}
where ${\cal L}_\eta$ denotes the Lie-derivative along $\eta_i$.
Further, a spacetime is said to admit a family of contracted
Ricci-collineation if
\begin{equation}
g^{ij}{\cal L}_\eta R_{ij}=0,
\end{equation}
which leads to the conservation law generator\footnote{Due to a
typographical error in [8], the same
definition of the conservation 
law generator for a contracted Ricci-collineation was also assigned
to the  Ricci-collineation. But this did not affect the
results of the paper as the Ricci-collineation was never used in the
paper.} 
\begin{equation} 
\left[T^j_m \eta^m + \left(\frac{\Lambda}{8\pi G} -
\frac{1}{2}T\right)\eta^j\right]_{;j}=0,
\end{equation}
if the Einstein field equations
\begin{equation}
R^{ij}-\frac{1}{2} R g^{ij}=-8\pi G\left(T^{ij}-\frac{\Lambda}{8\pi
G}g^{ij}\right)
\end{equation}
are satisfied. We consider units with $c=1$.
Recalling that the energy density associated with vacuum is
$\rho_{\mbox{{\scriptsize v}}} = \Lambda/8\pi G$ with its pressure 
$p_{\mbox{{\scriptsize v}}}=-\rho_{\mbox{{\scriptsize v}}}$ implying
$T^{ij} _{\mbox{{\scriptsize v}}} \equiv-(\Lambda/8\pi G) g^{ij}$, the quantity
appearing in the parentheses
on the right hand side of equation (4) can be written as 
\begin{equation}
T^{ij} _{\mbox{{\scriptsize tot}}} \equiv T^{ij} + 
T^{ij} _{\mbox{{\scriptsize v}}} =(\rho _{\mbox{{\scriptsize tot}}}
+p _{\mbox{{\scriptsize tot}}})u^iu^j + p _{\mbox{{\scriptsize tot}}}g^{ij},
\end{equation}
which represents the energy momentum tensor of the total matter, i.e.,
ordinary matter plus vacuum. Here $\rho _{\mbox{{\scriptsize tot}}}=\rho+
\rho_{\mbox{{\scriptsize v}}}$,
$p _{\mbox{{\scriptsize tot}}}=p+p_{\mbox{{\scriptsize v}}}=p-
\rho_{\mbox{{\scriptsize v}}}$ and $u^i$ is the
normalized velocity 4-vector. The Bianchi identities then confirm
the conservation of $T^{ij} _{\mbox{{\scriptsize tot}}}$ and not the 
conservation of $T^{ij}$ and $T^{ij} _{\mbox{{\scriptsize v}}}$ separately
unless $\Lambda=$ constant.

If we consider $\eta^i \propto u^i$, equation (3) reduces to 
\begin{equation}
\{(\rho _{\mbox{{\scriptsize tot}}} + 3p _{\mbox{{\scriptsize tot}}})u^j\}_{;j}=0,
\end{equation}
which may be interpreted as the conservation of generalized {\it momentum
density}. This is an important result in its own right , but it implies even
more. To understand the full meaning of this conservation law and compare
it with the existing results, let us consider the Robertson-Walker metric
\begin{equation} 
ds^2=-dt^2 + S^2(t) \left\{\frac{dr^2}{1-kr^2} +
r^2(d\theta^2+\sin^2\theta d\phi^2)\right\},
\end{equation}
characterized by the scale factor $S(t)$ and the curvature index
$k\in \{-1,0,1\}$ of the spatial hypersurfaces $t=$ constant.
Now the Einstein field equations (4) yield two independent equations:

\noindent
the Raychaudhuri  equation:

\begin{equation}
-\frac{\ddot S}{S}=\frac{4\pi
G}{3}\left(\rho _{\mbox{{\scriptsize tot}}}+3p _{\mbox{{\scriptsize tot}}}\right)
\end{equation}

\noindent
and the Friedmann equation:

\begin{equation}
\frac{\dot S^2}{S^2}+\frac{k}{S^2}=\frac{8\pi
G}{3}\rho _{\mbox{{\scriptsize tot}}}.
\end{equation}
By using (7), the conservation law (6) reduces to

\begin{equation}
(\rho _{\mbox{{\scriptsize tot}}} + 3p _{\mbox{{\scriptsize tot}}})S^3 =
\mbox{constant} = A ~ ~ \mbox{(say)},
\end{equation}
which is the central equation of our analysis and may be interpreted as 
the conservation of {\it total active gravitational
mass}, taken matter and vacuum together, of a comoving sphere of
 radius $S$. Note that $(\rho + 3p)$ is defined as the
{\it active gravitational mass density} of the universe [15]. Obviously
 the conservation law (10) reduces to the pressure-less
phase of the standard big bang (FLRW) model for $p=\Lambda=0$.
 
To understand the presence of pressure and vacuum terms in equation (10), we
consider the Raychaudhuri equation (8) which, being the analogue of Newtonian
gravitation, suggests that the gravitational attraction, experienced by
a unit test mass, is exerted in fact
not only by $\rho$ as in the Newtonian theory but rather by the {\it active
gravitational mass density} $(\rho+3p)$, which exhibits the relativistic
effects.  Associated with the attractive force 
$G {\cal A}/S^2$ (${\cal A} \equiv \frac{4\pi}{3}(\rho+3p)S^3$ being the 
{\it active gravitational mass} from the gravitating matter) in equation (8),
there is a repulsive force $-\Lambda S/3$ due to a positive $\Lambda$ (or an
additional  attraction  with a negative $\Lambda$).
It is this repulsive force which drives inflationary expansion in the early 
vacuum-dominated universe.  It is obvious that the constant $A$ 
(the {\it total active gravitational mass}) might take different values in
different phases of evolution depending upon the relative dominance of these
two terms of  opposite character in equation (8). This implies that the 
{\it total active gravitational mass} might be conserved phase-wise
only and not throughout the evolution. (For details, see [8]).

One may also
note that the left hand side of equation (8) is the Gaussian curvature of
the two-dimensional surface specified by varying $r$ and $t$, keeping
$\theta$ and $\phi$ constant, in equation (7) and may be considered as the
curvature of the homogeneous and isotropic spacetime. Thus  equation (8)
implies that the curvature of spacetime is governed by the total {\it active
gravitational mass density} of the universe.
In this view,  a naive assumption would be that the cause of curvature
of spacetime, i.e., the {\it total active gravitational mass} be
conserved, justifying our symmetry consideration leading to the conservation 
law (10). Equation (10) also implies, via equation (8), that
the curvature of spacetime
evolves as $S^{-3}$, which quickly transforms the spacetime from a
state of large curvature to a state of flatness.

It would be worth while to mention that if one wishes to describe the 
radiation-dominated universe in the Newtonian framework (extending
the work of McCrea and Milne), one needs to use ($\rho+3p$) instead of $\rho$
only in Poisson's equation to get the right answer. This is consistent with
our using ($\rho+3p$) for gravitational mass density.

Equations (8) and (10), taken together, yield

\begin{equation}
-\ddot S=\frac{4\pi G A}{3S^2},
\end{equation}
which integrates to

\begin{equation}
\dot S^2=\frac{8\pi GA}{3S}+B,                                             
\end{equation}
where $B$ is a constant of integration. This supplies the dynamics of the
scale factor. Equations (9), (10) and (12) may be used to obtain

\begin{equation}
\rho _{\mbox{{\scriptsize tot}}}=\frac{A}{S^3}+\frac{3(B+k)}{8\pi GS^2},
\end{equation}

\begin{equation}
p _{\mbox{{\scriptsize tot}}}=-\frac{(B+k)}{8\pi GS^2},
\end{equation}
which give the equation of state of the {\it perfect fluid} constituting
the
total matter of the universe as

\begin{equation}
p _{\mbox{{\scriptsize tot}}}^3=K(\rho _{\mbox{{\scriptsize tot}}}+3p _{
\mbox{{\scriptsize tot}}})^2,
\end{equation}
where $K=-(B+k)^3/(8\pi G)^3 A^2.$ This is a physically reasonable
equation of state since d$p_{\mbox{{\scriptsize tot}}}
/$d$\rho_{\mbox{{\scriptsize tot}}}=2p_{\mbox{{\scriptsize tot}}}/
3(\rho_{\mbox{{\scriptsize tot}}}+p_{\mbox{{\scriptsize tot}}})$ indicating
that d$p_{\mbox{{\scriptsize tot}}}$/d$\rho_{\mbox{{\scriptsize tot}}}
\leq 1/3$ for $\rho_{\mbox{{\scriptsize tot}}}\geq 
p_{\mbox{{\scriptsize tot}}}$ (provided $(\rho_{\mbox{{\scriptsize tot}}}+
p_{\mbox{{\scriptsize tot}}})>0)$. It may be noted that the
equation of state (15) breaks down for $p_{\mbox{{\scriptsize tot}}}=0$
(in the same way as does the
usual barotropic equation of state for a perfect fluid for $p=0$), in
which case
equations (13) and (14) may be treated as the parametric equations
of state. Consequences of the resulting models for the case $\Lambda=0$
have been discussed elsewhere by Abdussattar and Vishwakarma [9]  where the 
models obviously get constrained by  $A\geq 0$ and $(B+k)\leq 0$ and
by the additional constraint $S\leq 8\pi GA/3\mid B+k\mid$ when $(B+k)<0$.

When $\Lambda\neq0$, the equation of state (15) does not supply
information on the ordinary matter source. If the ordinary matter is
baryonic with its usual barotropic equation of state 

\begin{equation}
p=w\rho, ~ ~ ~ 0\leq w\leq 1,
\end{equation}
this simply implies that $\Lambda$ cannot just remain constant. The reason is
obvious. We now have 4 independent equations (12)-(14) and (16) in 3
unknowns $S$, $\rho$ and $p$.  This over-determinacy can be compensated 
by allowing at least one of the remaining parameters to vary.
The only such parameter of interest is $\Lambda$ if we keep $G$ as a 
constant (the cases with variable $G$ have been
discussed elsewhere [10]). 
It is obvious that we would have reached the same conclusion, had
we considered any other assumption in place of (2). However, this symmetry, 
as we have seen, has its own significance. Equations (13), (14) and
(16) thus yield

\begin{equation}
\Lambda=\frac{8\pi
G}{(1+w)}\left[\frac{w 
A}{S^3}+\frac{(1+3w)(B+k)}{8\pi G S^2}\right], 
\end{equation}

\begin{equation}
\rho=\frac{1}{(1+w)}\left[\frac{A}{S^3}+\frac{(B+k)}{4\pi G S^2}\right].                                             
\end{equation}

It may be mentioned that $\Lambda$ varying as $S^{-2}$, which the 
present model has in the pressure-less phase of evolution, has also been 
considered by several authors to explain the present small value of 
$\Lambda$ [11, 12]. The ansatz is primarily due to Chen and Wu [11] who 
postulated it through dimensional
arguments made in the spirit of quantum cosmology. Recently Jafarizadeh et
al [16] have calculated the tunneling rate with a
cosmological constant
decaying as $S^{-m}$ and concluded that the most probable cosmological
term with the highest tunneling rate occurred at $m=2$. However, the 
present model differs from the above-mentioned models in the sense that
contrary to the $\Lambda \sim S^{-2}$ throughout the evolution (as has 
been assumed in these models), $\Lambda$ varies 
differently in the different phases of evolution  in the present model
 as is clear from equation (17).

While comparing the model with SN Ia data, we shall be interested in the 
effects which occurred at redshift $<1$. We, therefore, neglect radiation
and consider $w=0$ in the following. (The parameters in the 
early phase of evolution can be calculated by following  [8]). 
With this in view, we recast equations (8) and 
(9) in the following forms to give the relative contributions of the
different cosmological parameters at the present epoch:

\begin{equation}
2[q_0 + \Omega_{\Lambda 0}]=\Omega_{0},
\end{equation}

\begin{equation}
1+\Omega_{k0} = \Omega_{0} + \Omega_{\Lambda 0},
\end{equation}
where $\Omega\equiv 8\pi G\rho/3H^2$, $\Omega_k\equiv k/S^2H^2$
and $\Omega_{\Lambda}\equiv\Lambda/3H^2$ are,
respectively, the dimensionless forms of the density, the curvature
and the cosmological constant parameters. and the subscript
$0$ characterizes the value of the quantity at the present epoch. 
Equations (10) and (12) may be used to give the values of the constants
$A$ and $B$ as

\begin{equation}
A=(\Omega_0-2\Omega_{\Lambda 0}) \frac{3S_0^3 H_0^2}{8\pi G},
\end{equation}
 
\begin{equation}
B=(2\Omega_{\Lambda0}-\Omega_0+1)H_0^2 S_0^2.
\end{equation}

\vspace{1cm}

\noindent
{\bf 3. MAGNITUDE-REDSHIFT RELATION IN THE MODEL}

\vspace{.2cm}

The cosmic distance measures, like the {\it luminosity
distance} and the {\it angular size distance}, depend
sensitively on the spatial curvature and the expansion dynamics
of the models and consequently on the present densities of the various
energy components and their equations of state.
For this reason, the magnitude-redshift relation for distant
{\it standard candles}  and the angular size-redshift relation for 
distant {\it standard measuring rods} have been proposed as potential
tests for cosmological models and play crucial role in determining
cosmological parameters.

Let us consider that the observer at $r=0$ and $t=t_0$ receives light
emitted at $t=t_1$ from a source of absolute luminosity $L$
located at a radial distance $r_1$. The cosmological redshift $z$
of the source is related with $t_1$ and $t_0$
by $1+z=S(t_0)/S(t_1)$. If the (apparent) luminosity of the source
measured by the observer is $l$,
the {\it luminosity distance} $d _{\mbox{{\scriptsize L}}}$ of the
source, defined by

\begin{equation}
l \equiv \frac{L}{4\pi d _{\mbox{{\scriptsize L}}}^2},
\end{equation}
is then given by

\begin{equation}
d _{\mbox{{\scriptsize L}}}=(1+z) S_0 ~ r_1.
\end{equation}
For historical reasons, the observed and absolute 
luminosities $l$ and $L$ are defined, respectively, in terms of
the K-corrected observed and absolute magnitudes $m$ and $M$ as
$l=10^{-2m/5}\times 2.52 \times 10^{-5}$ erg cm$^{-2}$ s$^{-1}$ and 
$L=10^{-2M/5}\times 3.02 \times 10^{35}$ erg s$^{-1}$  [17]. When
written in terms of $m$ and $M$,  equation (23) yields

\begin{equation}
m(z;{\cal M}, \Omega_0,\Omega_{\Lambda 0})={\cal M} +
5 \mbox{log}_{10}\{ {\cal D} _
{\mbox{{\scriptsize L}}}(z; \Omega_0,\Omega_{\Lambda 0})
\},
\end{equation}
where ${\cal M}=M-5\mbox{log}_{10}H_0 +25$ and ${\cal D}_{
\mbox{{\scriptsize L}}}(z; \Omega_0,\Omega_{\Lambda 0})
\equiv H_0 ~ d _{\mbox{{\scriptsize L}}}(z; \Omega_0,\Omega_{\Lambda 0}, H_0)$
is the dimensionless {\it luminosity distance}.
Here $d _{\mbox{{\scriptsize L}}}$ is measured in Mpc. By using equation (7), 
the coordinate distance $r_1$, appearing in equation (24), yields

\begin{equation}
\psi (r_1) =\int_{S_0/(1+z)}^{S_0} \frac{dS}{S\dot S}
\end{equation}
with

\[
\psi(r_1)=\sin^{-1} r_1, ~ ~ ~ ~ k=1
\]
\[~ ~ ~ ~  ~ ~ ~ =r_1, ~ ~ ~  ~ ~  ~ ~ ~ ~ ~ ~ ~ k=0
\]
\begin{equation}
~ ~ ~ ~ ~ ~ ~ ~ ~ ~ ~ =\sinh^{-1} r_1, ~ ~ ~ k=-1.
\end{equation}  
By the use of (12), (21) and (22), equation (26) yields

\begin{equation}
\psi(r_1) = \frac{1}{S_0 H_0} \int_0^z
[(2\Omega_{\Lambda0}-\Omega_0 + 1)(1+z')^2
-(2\Omega_{\Lambda0}-\Omega_0)(1+z')^3 ]^{-1/2} dz'.
\end{equation}  
Equations (24), (27) and (28) can also be combined into a single
compact equation as

\vspace{.4cm}
\noindent
${\cal D} _{\mbox{{\scriptsize L}}}(z; \Omega_0,\Omega_{\Lambda 0})=$
\begin{equation}
\frac{(1+z)}{\sqrt{{\cal K}}} ~ \xi\left(\sqrt{{\cal K}} ~ \int_0^z
[(2\Omega_{\Lambda0}-\Omega_0 + 1)(1+z')^2
-(2\Omega_{\Lambda0}-\Omega_0)(1+z')^3 ]^{-1/2} dz'\right),
\end{equation}
where

$\xi(x)=\sin (x)$ ~ ~ with ~ ~ ${\cal K}=\Omega_{k0}$ ~ ~ ~when ~ ~
$\Omega_{k0}>0$,

$\xi(x)=\sinh (x)$  ~ ~with ~ ~${\cal K}=-\Omega_{k0}$~ ~ when ~ ~
$\Omega_{k0}<0$,

$\xi(x)=x$ ~ ~ ~ ~ ~ with ~ ~ ~${\cal K}=1$ ~ ~ ~ ~when ~ ~ 
$\Omega_{k0}=0$.

Thus for given ${\cal M}$, $\Omega_0$ and $\Omega_{\Lambda0}$, equations
(25) and (29) give the predicted value of $m(z)$ at a given $z$. By using
the K-corrected effective magnitudes $m_i^{\mbox{{\scriptsize eff}}}$, 
which have also been corrected for the light-curve width-luminosity relation and
galactic extinction, and  using the same standard errors $\sigma_{z,i}$ and 
$\sigma_{m_i^{\mbox{{\scriptsize eff}}}}$ of the $i$th supernova with
redshift $z_i$ as used by Perlmutter et al, we compute $\chi^2$ according to

\begin{equation} 
\chi^2=\sum_i \frac{[m_i^{\mbox{{\scriptsize eff}}} - m(z_i)]^2}
{(\sigma_{z,i}^2 + \sigma_{m_i^{\mbox{{\scriptsize eff}}}}^2)}.
\end{equation}
The best fit parameters are obtained by minimising this equation.
We note that equation (29) is sensitive to $\Omega_0$ and $\Omega_{
\Lambda0}$ for distant sources only. For the nearby sources (in the
low-redshift limit), equations (25) and (29) reduce to

\begin{equation}
m(z)= {\cal M} + 5 ~ log_{10}z,
\end{equation}
which can be used to measure ${\cal M}$ by using low-redshift 
supernovae-measurements
that are far enough into the Hubble flow so that their peculiar
velocities do not contribute significantly to their redshifts.

\vspace{1cm}

\noindent
{\bf 4. NUMERICAL RESULTS}

\vspace{.2cm}
We consider the low-redshift data on $m^{\mbox{{\scriptsize eff}}}$ and $z$
from the Calan-Tololo sample of 16 supernovae (excluding 2 outliers from the 
full sample of 18 supernovae) to estimate ${\cal M}$ from equations [30] and
(31).  This gives ${\cal M}=24.03$ (in units with $c=1$). This value is used in
equation (25) to estimate $\Omega_0$ and $\Omega_{\Lambda0}$ from the 
high-redshift measurements.
For this purpose we consider the data set on $m^{\mbox{{\scriptsize eff}}}$ 
and $z$ of 38 supernovae from the Supernova Cosmology Project (excluding 2 
outliers and 2 likely reddened ones from the full sample of 42 supernovae).
We perform a two-parameter fit of this data set by following the fitting 
procedure of Perlmutter et al\footnote{
It may be mentioned that Perlmutter et al [2] have also fitted the data for
only 3 parameters ${\cal M}$, $\Omega_0$ and $\Omega_{\Lambda0}$ and
not for the 4 parameters $\alpha$, ${\cal M}$, $\Omega_0$
and $\Omega_{\Lambda0}$ as mentioned in their paper (from
a personal discussion with Professor R. S. Ellis, one of the authors
of the paper). A self-consistent 4-parameter fit has been done by
Efstathiou et al[18].}
considered in their Fit M (see Fig. 5 (f) in their paper [2]).

We find that the data shows a very good fit to the model giving the 
best-fitting
flat model ($\Omega+\Omega_{\Lambda}=1$) as $\Omega_0=0.54$ with
$\chi^2=45.34$, which shows that the fit is almost as good as to the
constant $\Lambda$-flat FLRW model: $\Omega_0=0.40$ with $\chi^2=44.92$.

The minimisation process gives the global best-fitting solution 
(calculated by giving free rein to $\Omega_0$ and $\Omega_{\Lambda0}$) as 
$\Omega_0=1.76$ and $\Omega_{\Lambda0}=1.34$ with
$\chi^2=44.78$ which shows a rather high value of $\Omega_0$ and does
not seem realistic in view of the small observed value of $\Omega_0$. 
One can see that the data  predicts
an accelerating expansion ($q_0<0$) of the universe as in the
constant $\Lambda$-FLRW model.

The fit of the flat model to the actual data points
has been shown in Figure 1, where we have compared it with the simplest
$\Lambda=0$ model, i.e., the Einstein-de Sitter model ($\Omega=1$),
which is ruled out by the data ($\chi^2=115.81$). 

It may be mentioned that elsewhere [7], we have used this data in the model,
with a different fitting procedure, by fitting the low- and high-redshift
measurements simultaneously to equation (30). The best-fitting solutions so
obtained are in good agreement with those obtained here.

It would also be worth while to mention that the present model is consistent
not only with the supernovae-data but it also fits very well the data
on the angular size and redshift of the ultracompact radio sources
complied by Jackson and Dodgson [19] as well as the updated and modified
data on compact radio sources from Gurvits et al [7].

\centerline{{\epsfxsize=14cm {\epsfbox[50 250 550 550]{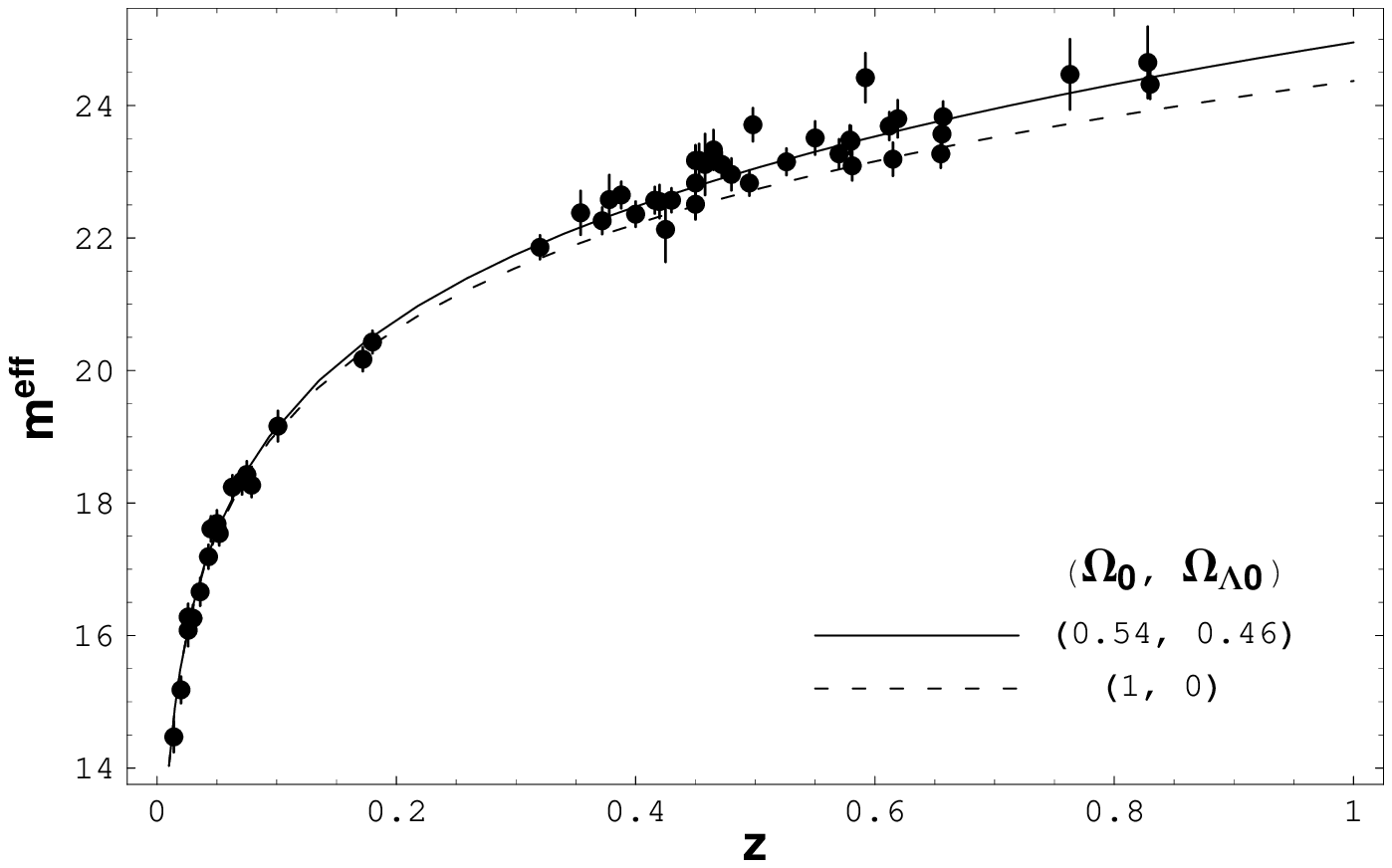}}}}

{\bf Figure 1.} Hubble diagram for 38 high-redshift and 16 low-redshift su-

pernovae: The solid curve represents the best-fitting flat model
($\Omega_0=$

0.54, $\Omega_{\Lambda0}=0.46$). For comparison, the canonical
Einstein-de Sitter model 

($\Omega_{k}=0$, $\Omega_{\Lambda}=0$) has also been plotted (dashed curve).

\vspace{1cm}

\noindent
{\bf 4. CONCLUSION}

\vspace{.2cm}
In order to solve the cosmological constant problem, which has been made even 
more acute by the consequences of the current observations of CMB and type Ia 
supernovae, several authors
have invoked a variable cosmological term $\Lambda(t)$. Instead of considering 
any ad-hoc assumption
for the variation of $\Lambda$, as has been mainly done by the authors, we
have considered a particular symmetry of the homogeneous, isotropic spacetime
which is the contracted Ricci-collineation along the fluid flow.
It have been shown that the incorporation of this additional symmetry into the 
RW model filled with
baryonic matter renders the cosmological term $\Lambda$ a decaying function 
of the scale factor. This helps in solving the cosmological constant problem.
This new symmetry in the model leads to the conservation of the 
{\it total active gravitational mass} of the universe.

The resulting model fits the SN Ia data from Perlmutter et al very well
and requires an accelerated expansion of the universe with a non-zero positive 
cosmological term. The goodness of fit of the data to the model
is almost the same as in the case of the constant $\Lambda$-FLRW model.
The estimates of the parameters for the best-fitting flat model are obtained as 
$\Omega_0=0.54$ and $\Omega_{\Lambda0}=0.46$. However, the global best-fitting
solution, $\Omega_0=1.76$ with $\Omega_{\Lambda0}=1.34$, does not seem
realistic (as is the case with the constant $\Lambda$-FLRW model; see
reference [2]) in  view of the small observed value of $\Omega_0$. It is also
noted that  the estimates of the density parameter $\Omega_0$ for the present
variable  $\Lambda$-model are found a bit higher than those for the constant 
$\Lambda$-FLRW model.

\vspace{.6cm}

\noindent
{\bf ACKNOWLEDGEMENTS}

\vspace{.2cm}
The author thanks the Department of Atomic energy, India for providing the
Homi Bhabha postdoctoral fellowship and the IUCAA for hospitality.

\vspace{1cm}

\noindent
{\bf REFERENCES}

\vspace{.2cm}

\noindent
1. deBernardis, P., et al (2000) Nature {\bf 404}, 955;
   Hanany, S., et al (2000) 

   ~Astrophys. J. {\bf 545}, L5.\\
2. Perlmutter, S., et al (1999) Astrophys. J. {\bf 517}, 565.\\
3. Riess, A. G., et al (1998) Astron. J.  {\bf 116}, 1009.\\
4. Chiba, M. and Yoshi, Y. preprint (astro-ph/9808321).\\
5. Sahni V and Starobinsky A (2000) Int. J. Mod. Phys. D {\bf 9}, 373; and
   the 
   
   ~references therein.\\
6. Garnavich et al (1998) Astrophys. J. {\bf 509}, 74.\\
7. Vishwakarma, R. G. (2001)  Class. Quantum Grav. {\bf 18}, 1159.\\
8. Abdussattar and Vishwakarma, R. G. (1996) Pramana-J. Phys. {\bf 47}, 41.\\
9. Abdussattar and Vishwakarma, R. G. (1995) Curr. Sci. {\bf 69}, 924.\\
10. Abdussattar and Vishwakarma, R. G. (1996) Indian J. Phys. B {\bf 70}, 321;
    
    ~(1997) Austral. J. Phys. {\bf 50}, 893;
    Vishwakarma, R. G. and Beesham, A. 
    
    ~(1999) IL Nuovo Cimento B {\bf 114}, 631.\\
11. Chen, W. and Wu, Y. S. (1990) Phys. Rev. D {\bf 41}, 695.\\
12. Abdel-Rahaman, A-M. M. (1992) Phys. Rev. D {\bf 45}, 3492;
   Carvalho, J. 
   
   ~C., Lima, J. A. S. and Waga, I. (1992) Phys. Rev. D {\bf 46}, 2404;
   Silveira, 
   
   ~V. and Waga I. (1994) Phys. Rev. D {\bf 50}, 4890 ;
   Waga, I. (1993) Astro-
   
   ~phys. J. {\bf 414}, 436.\\
13. Freese, K., Adams, F. C., Friemann, J. A. and Mottolla E. (1987) 

    ~Nucl. Phys. B, {\bf 287}, 797;
   Gariel, J. and Le Denmat, G. (1999) 
Class. 

~Quantum Grav. {\bf 16}, 149;
  Gasperini, M. (1987) Phys. Lett. B, {\bf 194}, 347;
  
~(1988) 
  Class. Quantum Grav. {\bf 5}, 521;
  Ozer, M. and Taha, M. O. (1987) 

~Nucl. Phys. B, {\bf 287}, 776;
  Peebles, P. J. E. and Ratra, B. (1988) Astro-

~phys. J. 
  {\bf 325}, L17;
  Vishwakarma, R. G. (2000)  Class. Quantum Grav. 

~{\bf 17}, 3833.\\
14. Collinson, C. D. (1970) Gen. Relativ. Grav. {\bf 1}, 137;
    Davis, W. R., 

    ~Green, L. H. and 
    Norris, L. K. (1976) Nuovo Cimento B {\bf 34}, 256.\\
15. Ellis, G. F. R. (1971)  in General Relativity and Cosmology, ed.
    Sacks 
    
    ~R. K., Academic Press.\\
16. Jafarizadeh, M. A., Darabi, F., Rezaei-Aghdam, A. and Rastegar A. R. 

    ~(1999) Phys. Rev. D {\bf 60}, 063514.\\
17. Weinberg, S. (1972) Gravitation and Cosmology, John Wiley.\\
18. Efstathiou, G., Bridle, S. L., Lasenby, A. N., Hobson, M. P. and Ellis, 
   
    ~R. S. (1999) MNRAS, {\bf 303}, L47.\\
19. Vishwakarma, R. G. (1999) preprint (gr-qc/9912106).\\

\end{document}